# Fundamental Science and Improvement of the Quality of Life-Space Quantization to MRI

M. J. Tannenbaum[1]
Brookhaven National Laboratory
Upton, NY 11973 USA

[1] Supported by U.S. Department of Energy, DE-AC02-98CH10886.

# 1. Science versus Technology – a false dichotomy

• Science is the study of the laws of nature and the properties of natural objects. It answers a fundamental need of human nature, the desire to understand.

• Technology is the application of scientific knowledge to make devices.

• Science improves with improving technology and vice versa.

• Many of the problems facing us require scientific discovery as well as technological development.

# 2. Scientific Discovery is vital for future progress—

This is generally believed by government and private industry in the U.S. Here are a few examples.

## 2.1 M.I.Pupin-Serbian immigrant and Columbia College graduate 1883

PhD Berlin 1889-Prof. Columbia 1889-1931-made significant contributions to long-range telegraphy and telephony-and a sizeable fortune. He attributed his success in part to a fortunate encounter with classical academics/mechanics. Inspired by this experience, he preached the study and promotion of “pure science”, which he called the “goose that laid the golden egg”.

Donated his estate to “pure science” – Pupin Laboratory at Columbia. [1]

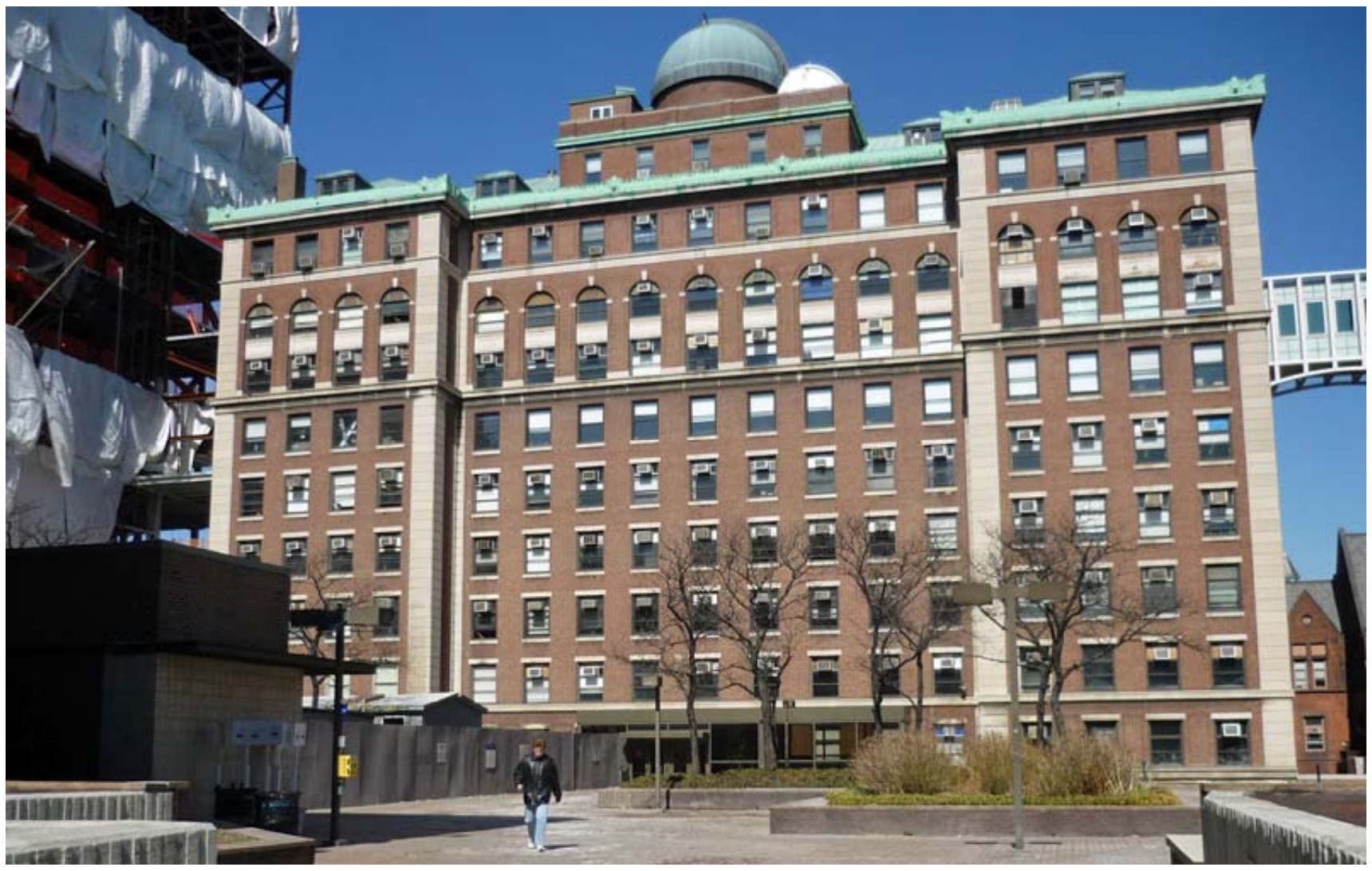

Fig. 1 Pupin Physics Laboratories, Columbia University, New York, NY, USA

### 2.2 G.A. Keyworth, science advisor to President R. Reagan 1981-86

"No federal research dollars, on average, gain more fruitful rewards than do those relatively few committed to basic research, the search for pure knowledge." [2]

### 2.3 U. S. National Academy of Sciences Report 2007

From "Rising Above the Gathering Storm: Energizing and Employing America for a Brighter Economic Future", Committee on Prospering in the Global Economy of the 21st Century, N. Augustine (chair), National Academy Press, Washington DC. Chapter 6.1, "Sowing the Seeds through Science and Engineering Research"

**Recommendation B:** ***Sustain and strengthen the nation's traditional commitment to long-term basic research that has the potential to be transformational to maintain the flow of new ideas that fuel the economy, provide security, and enhance the quality of life.***

**Implementation Actions:**
**Action B-1: Increase the federal investment in long-term basic research by 10% a year over the 7 years,** through reallocation of existing funds2 or if necessary through the investment of new funds. Special attention should go to the physical sciences, engineering, mathematics, and information sciences and to Department of Defense (DOD) basic-research funding. This special attention does not mean that there should disinvestment in such important fields as the life sciences (which have seen growth in recent years) or the social sciences. A balanced research portfolio in all fields of science and engineering research is critical to US prosperity.

## 3 An Example: Space Quantization to Magnetic Resonance Imaging (MRI) – A timeline from 1911-1977

**1911- Rutherford-Discovers the Nucleus**. By scattering alpha particles(from radioactive decay), on gold foils he finds that the positive charge of matter is restricted to a volume with radius 10 fm ($10^{-14}$ m=0.00000000000001m), the nucleus. The negative charged electrons are at a much larger radius.

**1913-Bohr Theory of Hydrogen Spectrum-Quantized electron orbits.**
In the Bohr theory, electrons orbit the nucleus, like planets around the sun. Radiation is emitted when electrons fall from a higher orbit to a lower orbit. (Figure 2). This results in a series of characteristic discrete spectral lines emitted by the different elements. The spectrum of hydrogen (one electron orbiting about one proton) was the simplest. Bohr explained an empirical formula by Balmer for the spectral wavelengths of light emitted by hydrogen by asuming that the electron orbits were quantized. They could only take on certain values given by integer "quantum numbers".

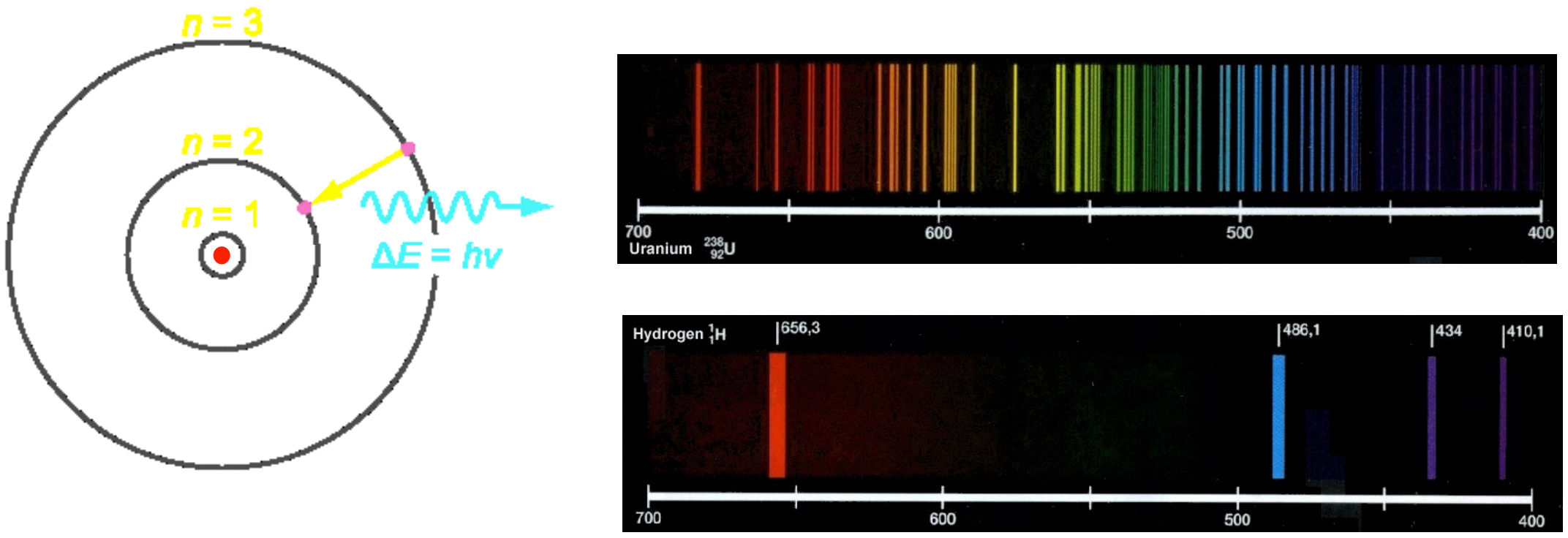

Figure 2. (Left) Schematic orbits of electrons with integer quantum numbers, n=1,2,3, about a nucleus. (Right-top) Emission spectrum of $_{92}U^{238}$, (Right-bottom) Emission spectrum of Hydrogen ($_1H^1$)

**1916-Sommerfeld-Bohr Model-Space Quantization-elliptical orbits**

Sommerfeld proposed that the electrons travel in elliptical rather than circular orbits. More important for the present discussion, in order to reproduce the observed spectral lines, Sommerfeld proposed that an electron orbit can not take any angle with respect to an external magnetic field, only integer projections of angular momentum. In classicsl physics, there is no such restriction—any angle is possible.

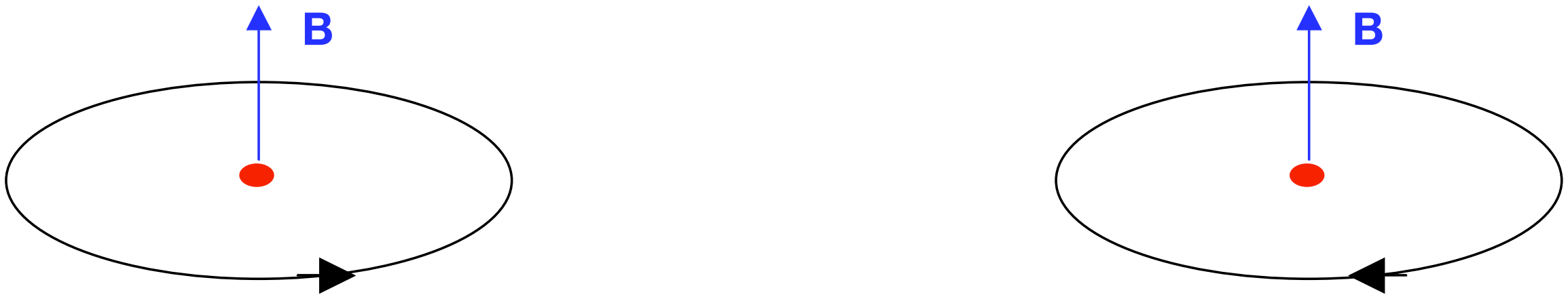

Figure 3. In an external magnetic field, B, the orbit of an electron with 1 unit of angular momentum can only take discrete orietations with respect to the magnetic field.

**1922-Stern Gerlach experiment proves space quantization**.

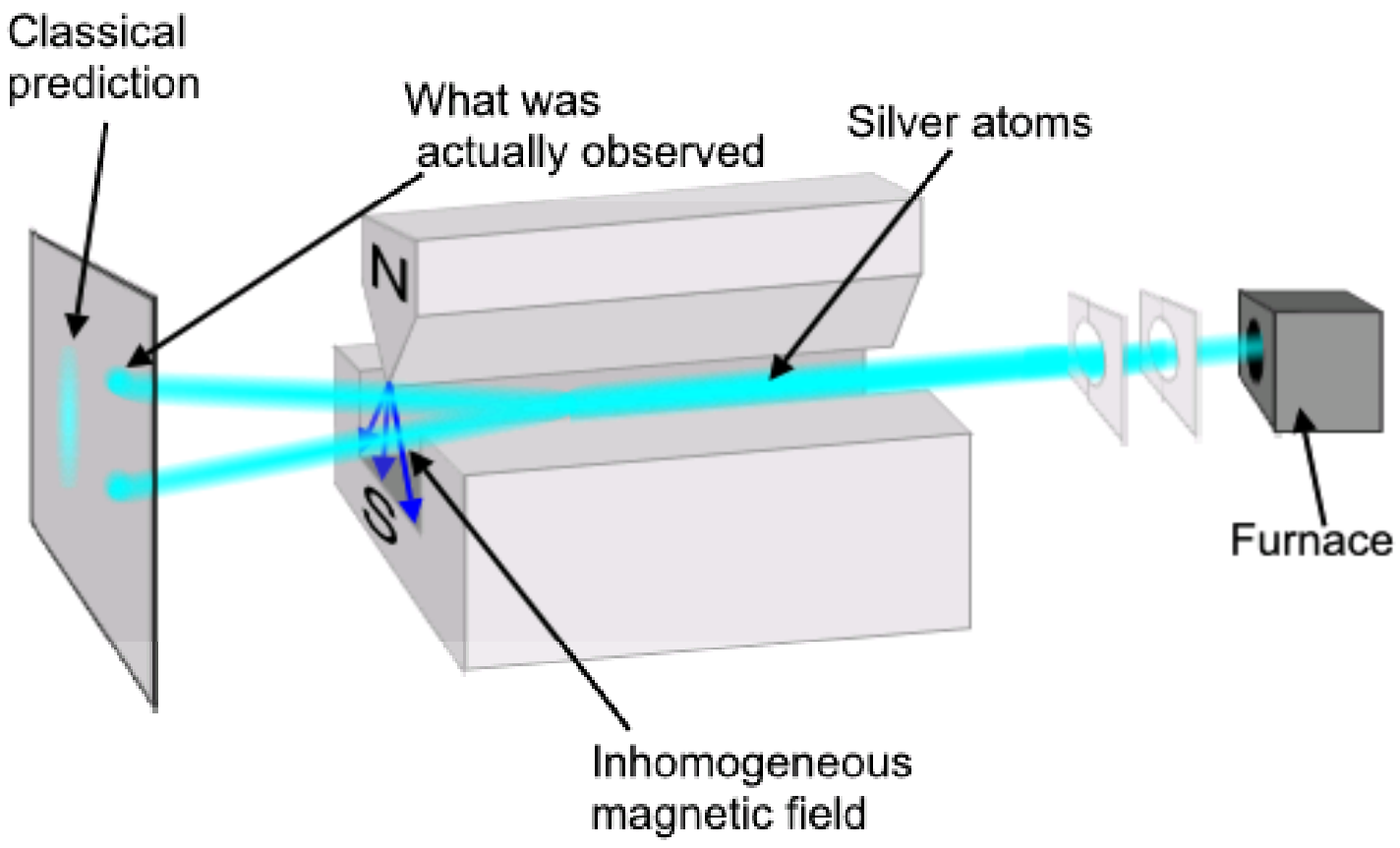

Figure 4. Stern Gerlach Atomic Beam experiment

Stern and Gerlach in Frankfurt created a beam of silver atoms and passed them through a non-uniform magnetic field, which deflects the beam proportional to the angular momentum projection along the B field. The beam doesn't spread, it splits only into 2 projections, proving space quantization.

**1925-Pauli trying to understand periodic table (2,8,18,32) proposes Exclusion Principle – "a** new quantum theoretic property of the electron, which I called a « two-valuedness not describable classically »"

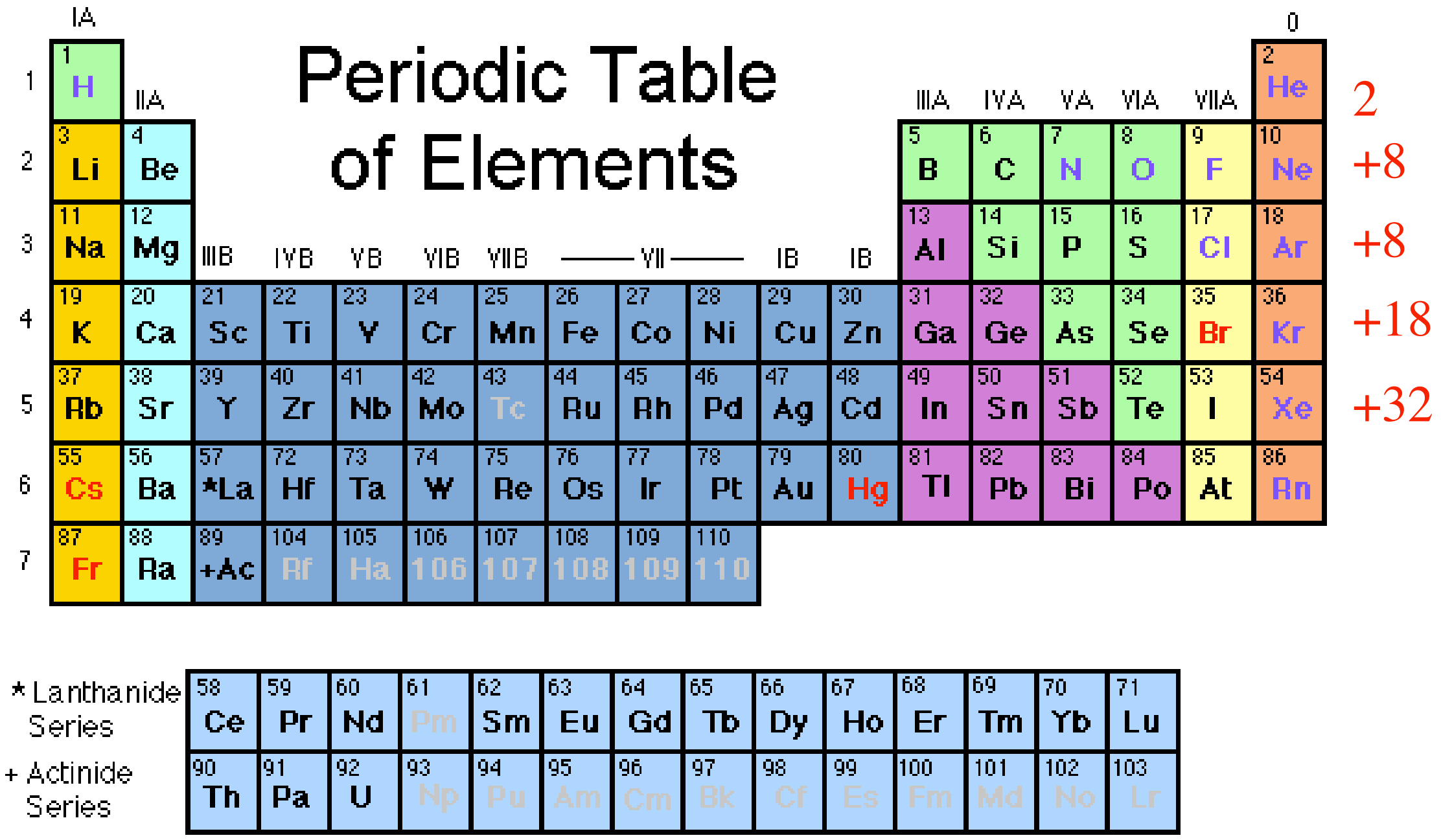

**1925--Goudsmit and Uhlenbeck** propose electron spin of 1/2 as the source of the two-valuedness. (two possibilities-spin ↑, spin ↓ )

**1927--I.I.Rabi receives PhD from Columbia**. He goes to Hamburg to work with Pauli. Instead, he learns atomic beams and does experiment with Stern.

**1929- Rabi returns to Columbia**, sets up Molecular beams Lab.

**1937-Rabi Invents "Molecular Beam Magnetic Resonance Method"**

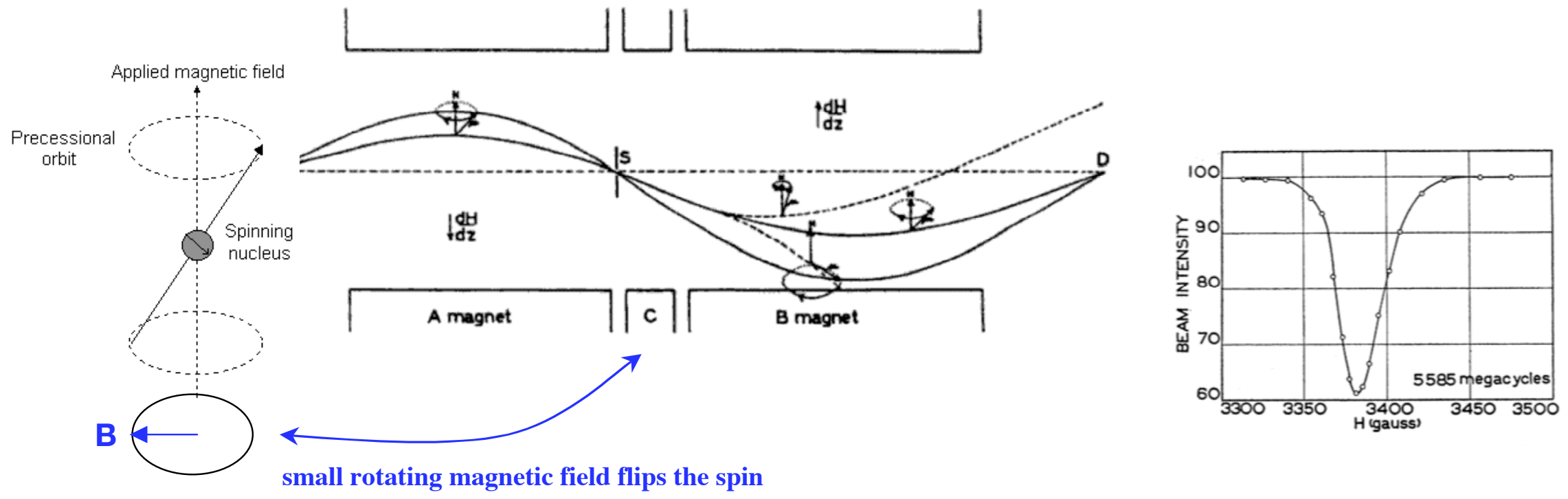

Figure 6. The molecular beam is deflected by the inhomogenious magnetic field in the left Stern-Gerlach apparatus (A). Rotating magnetic field at the resonant frequendy in the region c, flips the spin and deflects the beay away from axis [3]

The spin orieintation of a nucleus with a magnetic moment precessing around an external magnetic field is flipped by a rotating magnetic field at the “Larmor frequency”. In the two back to back Stern-Gerlach apparata, resonantly flipping the spin causes the beam to not return to the axis which causes a dip in the measured intensity.  The resonant displacement of the beam leads to measurement of magnetic moments of nuclei with high precision.

## 1939-1945 World II Intervenes-MIT Radiation Laboratory with I. I. Rabi as deputy director for scientific matters

MIT Rad. Lab. develops improved radio frequency sources and detectors  in additon to Radar, Loran, etc.

**1945-46—Nuclear Magnetic Resonance (NMR)** Purcell (paraffin) and Bloch (water) observe Nuclear Magnetic Resonance of protons in solid and liquid materials in an external magnetic field. (Protons have spin1/2 and act like little magnets.)

## Meanwhile, powerful magnets, computing technology, algorithms are developed leading to

**1971-1977 Magnetic Resonance Imaging (MRI)**—Damadian (SUNY-Brooklyn), Lauterbur (SUNY-Stony Brook), Mansfield ( Nottingham)

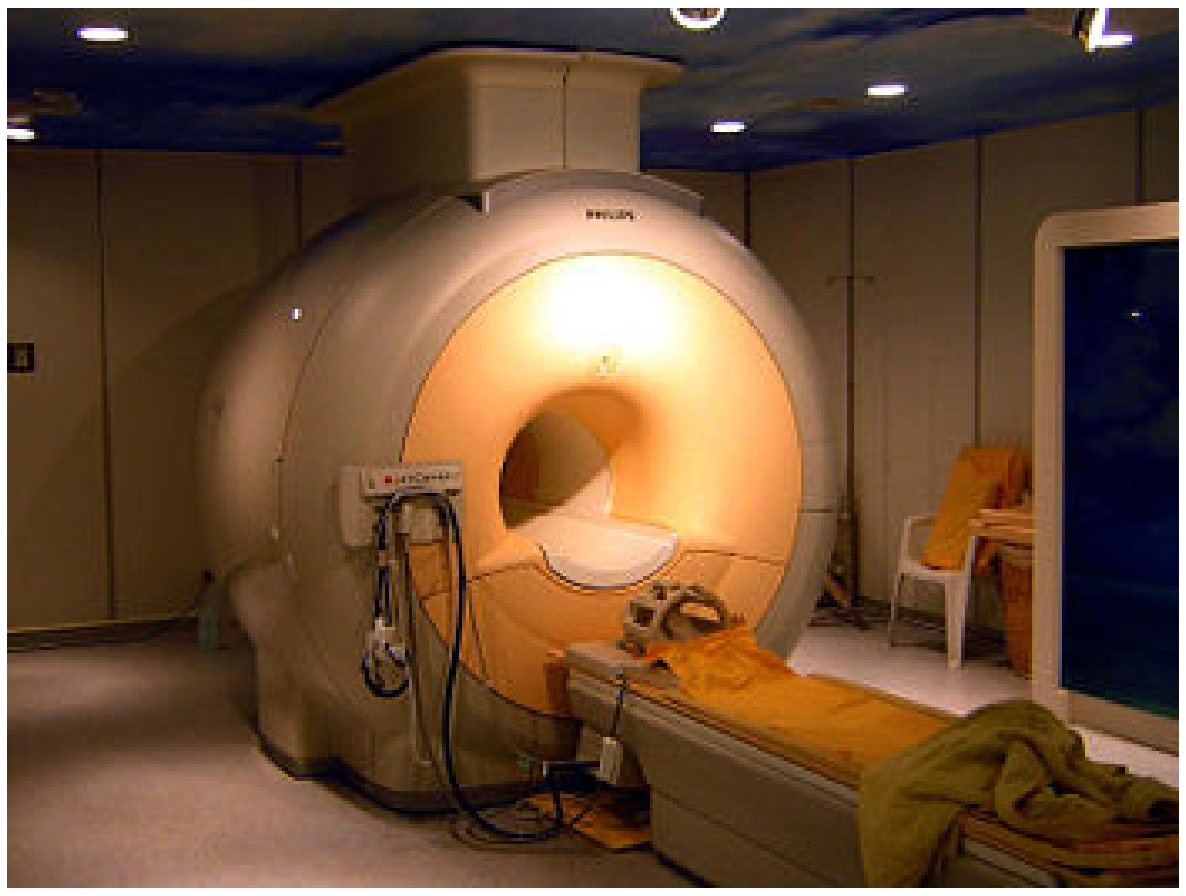

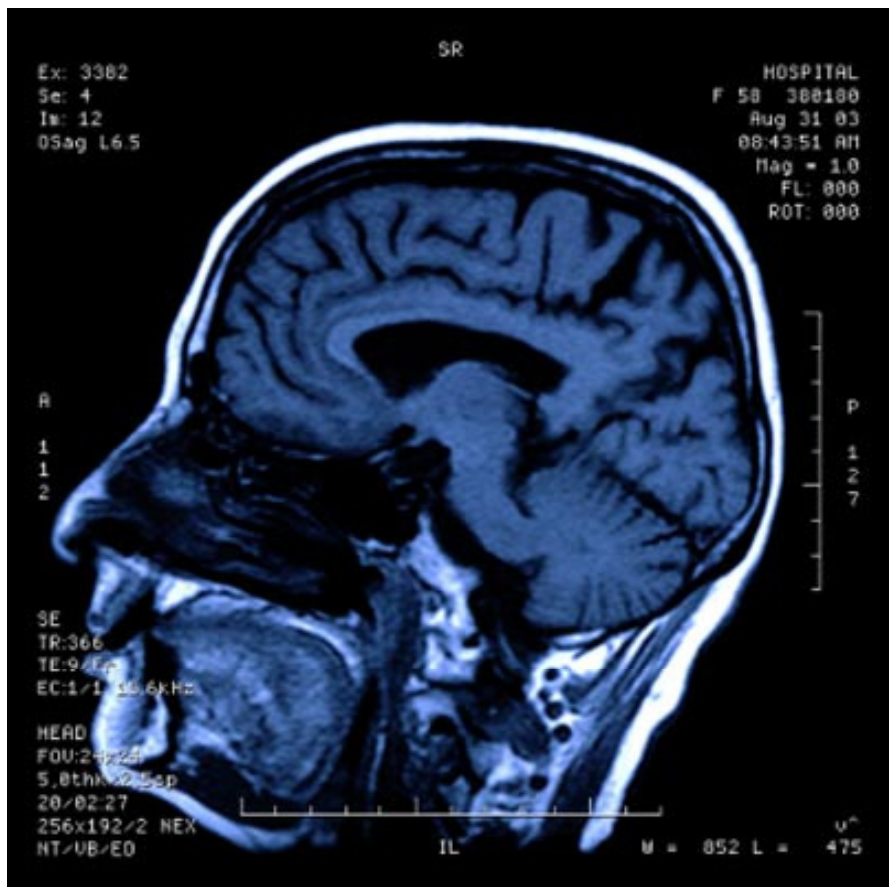

Figure 7. (Left) Modern MRI machine, a large electromagnet (axial magnetic field) with lots of fancy r.f. and computing power. (Right) Typical MRI picture of the head—examination of the brain and other delicate internal parts of the human body without surgery and with exquisite resolution.

**1967-Rabi retires from Columbia**-Photo of Rabi and Magnetic Resonance disciples

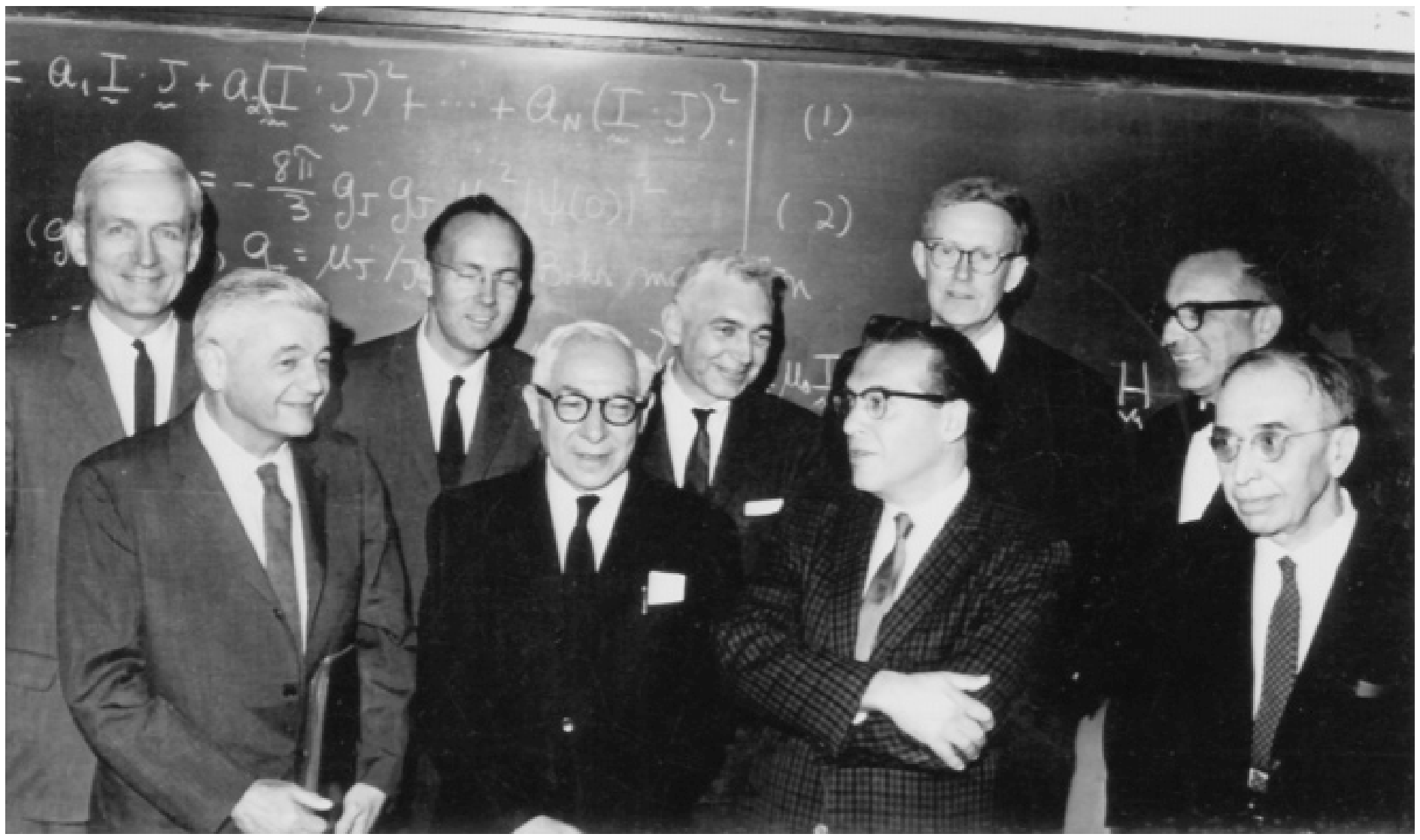

Figure 8. Rabi and disciples at the time of his retirement [4]. From left to right: N. Ramsey (Hydrogen Maser-most precise frequency source), G. Zacharias, C. Townes (Maser, leading to Laser), I.I. Rabi, V. Hughes (spin structure of nucleon), J. Schwinger (Quantum Electro Dynamics), E. Purcell (NMR), W. Nierenberg, G. Breit.

**Rabi was also instrumental in in creating both Brookhaven National Laboratory (U.S.) and CERN (Europe) for fundamental research.**

## 4 Modern Basic Research—what is inside the proton?

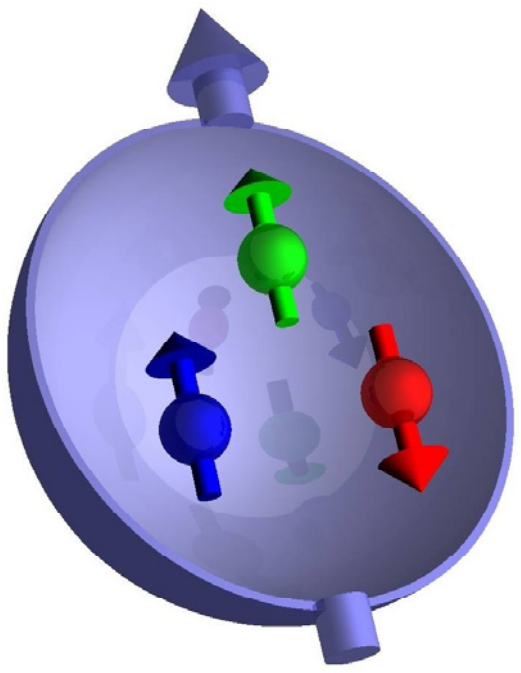

Protons (spin 1/2) are are composed of 3 quarks (spin 1/2). Quarks in protons come in 2 "flavors", up and down, and 3 "colors" RGB, where "color" and "flavor" are the modern quantum numbers.

We collide beams of polarized protons with each other and beams of polarized electrons with beams of polarized protons to determine where the mass and spin are located inside a proton.

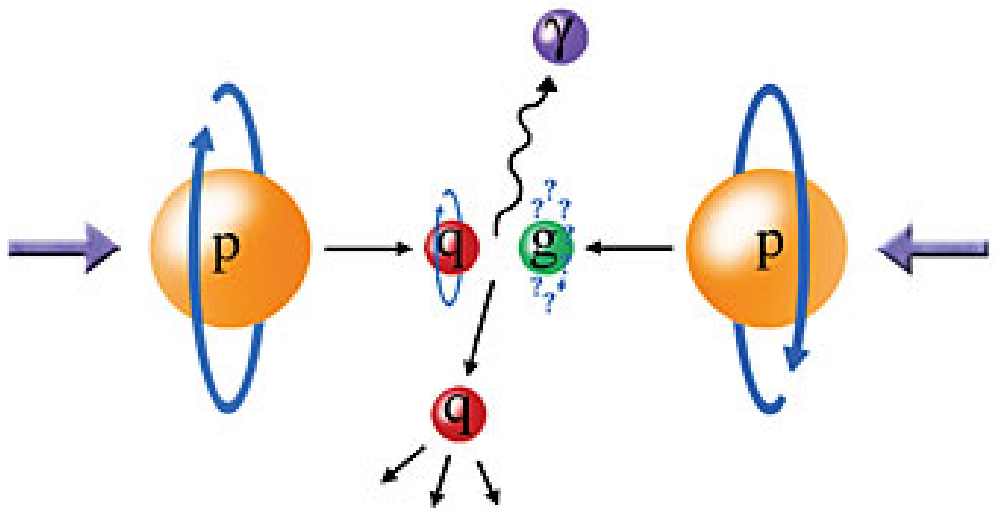

Like two Stern-Gerlace beams colliding with each other but with much higher energy.

## 4.1 Where is the spin located inside the proton?

Quarks only carry 1/2 the momentum of a moving proton: color-charged gluons (the quantum of the strong force) make up the rest.

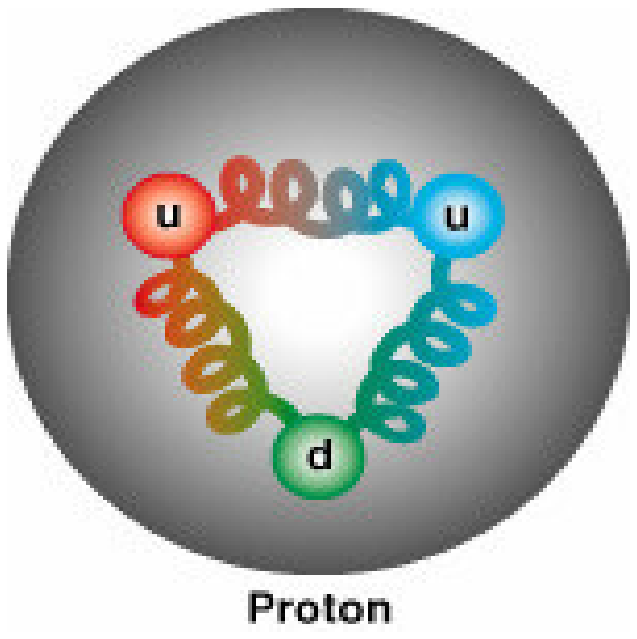

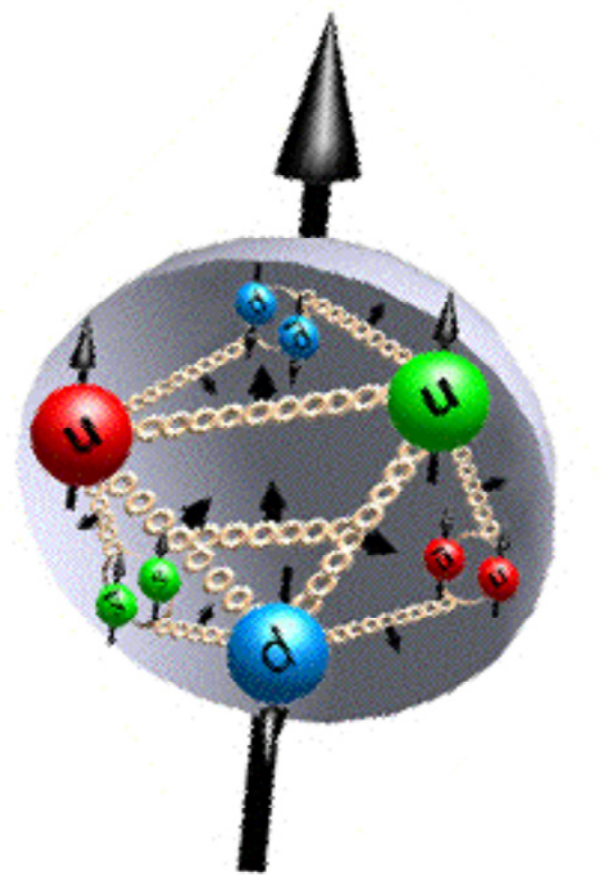

In addition to the 3 “valence” quarks and gluons, there are pair-produced sea quarks; and all these constituents of the proton have angular momentum. We haven’t yet figured out where the spin of the proton is.

However, we do know that 98% of the mass of the proton is due to the kinetic energy of the constituents, not to their rest masses!!

## 4.2 Where we do these experiments:RHIC—Relativistic Heavy Ion and polarized proton-proton collider on Long Island, visible from space.

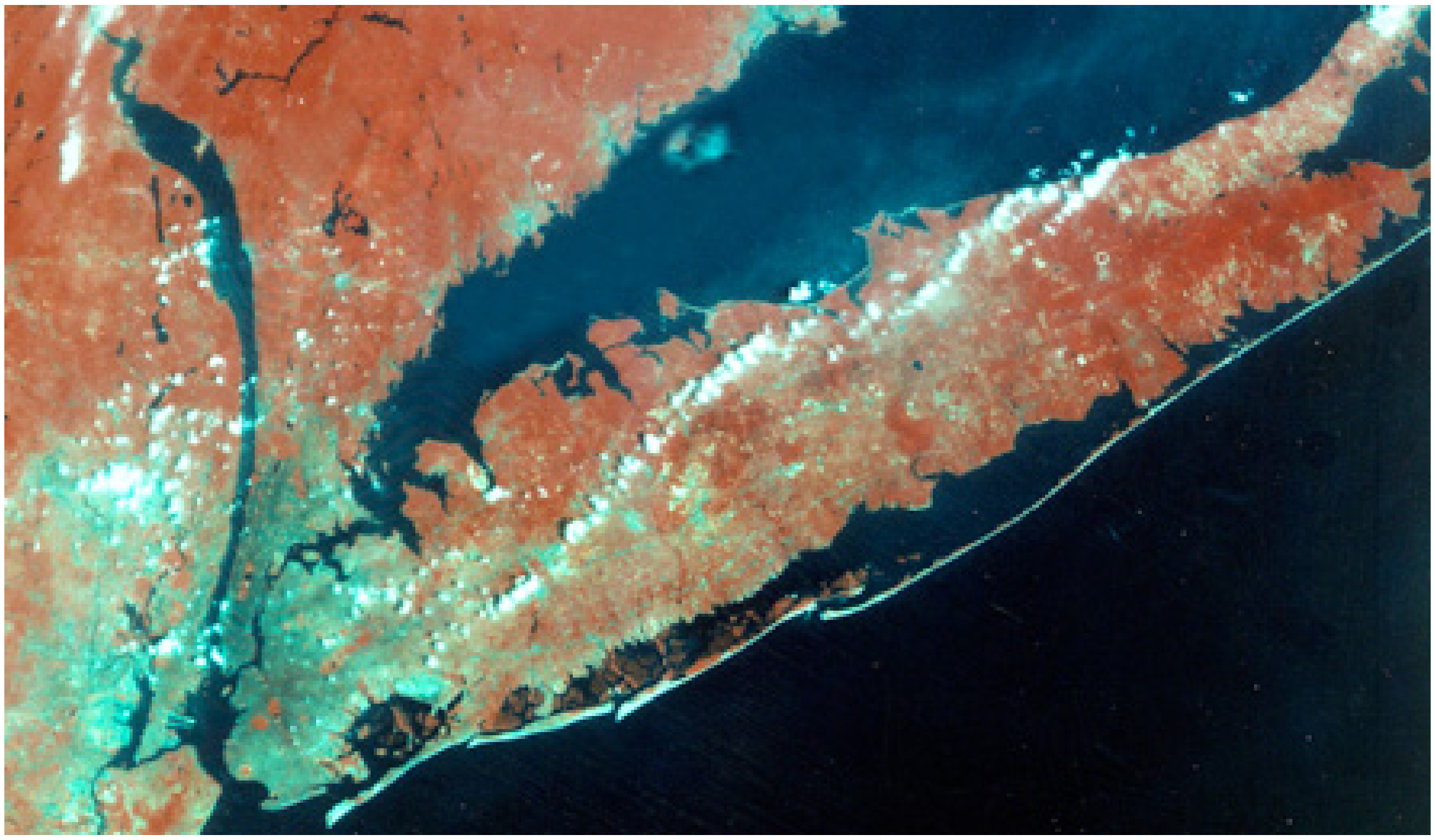

Figure 9. NASA Infra-red photo of New York Metro Region. RHIC is the white circle in the center of Long Island below the rightmost group of clouds. Manhattan island is clearly visible on the left side.

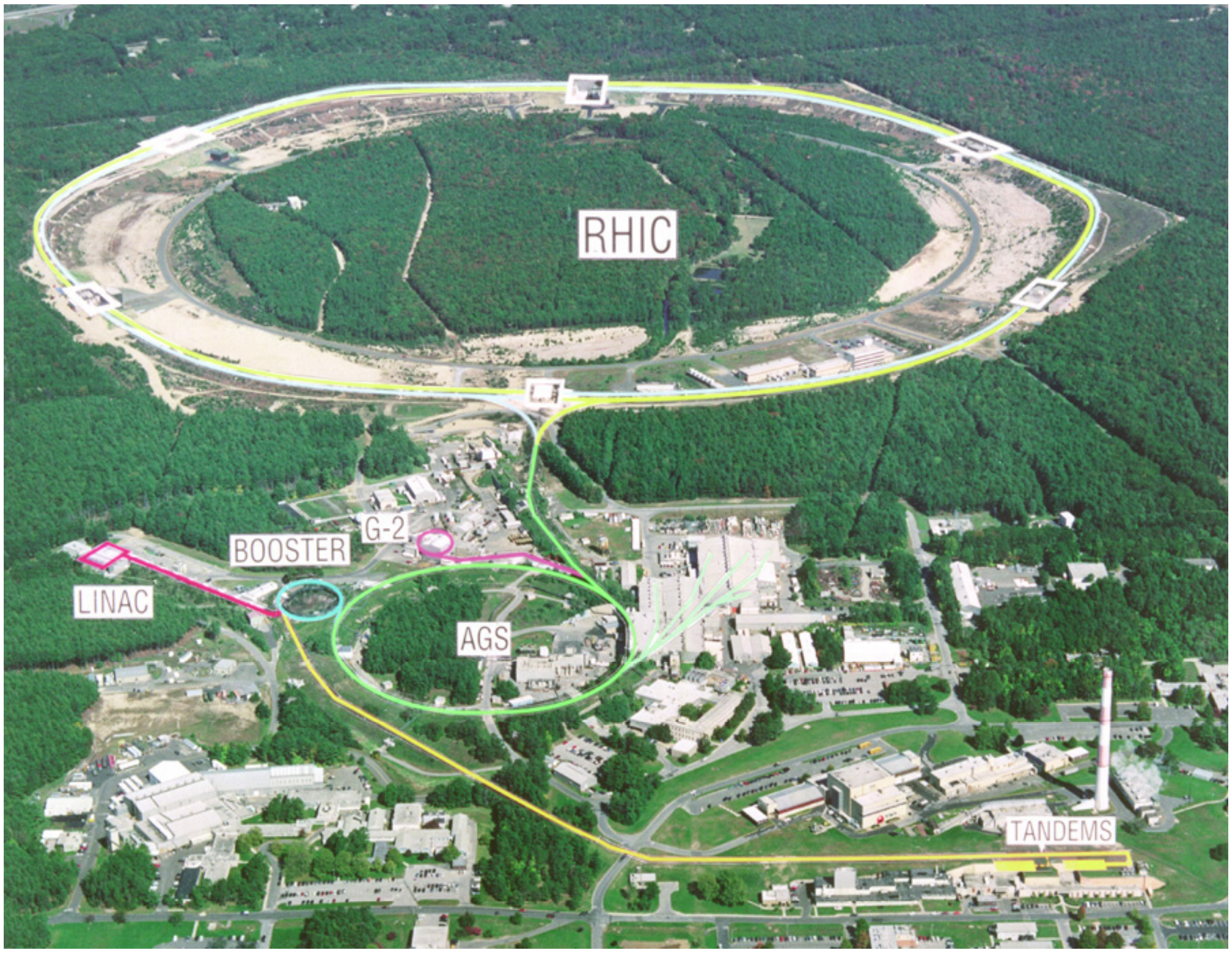

Figure 10. A closer view of RHIC at Brookhaven National Laboratory.The large circle without tree cover is excavation related to the tunnel containing the RHIC machine. The colored lines show the Linac, Booster accelerator for polarized proton injection, the tandem van de graaf accelerator and transfer line to the booster, and the AGS which accelerates the beams to an energy of 22 GeV per nucleon × Z/A where Z and A are the atomic number and weight of the nucleus.

**2000-RHIC begins operation:** an accelerator made entirely from superconducting magnets**. Superconductivity is another triumph of basic research—**Discovered by Kamerlingh-Onnes in 1911; but only in the 1970's were practical large superconducting magnets developed.

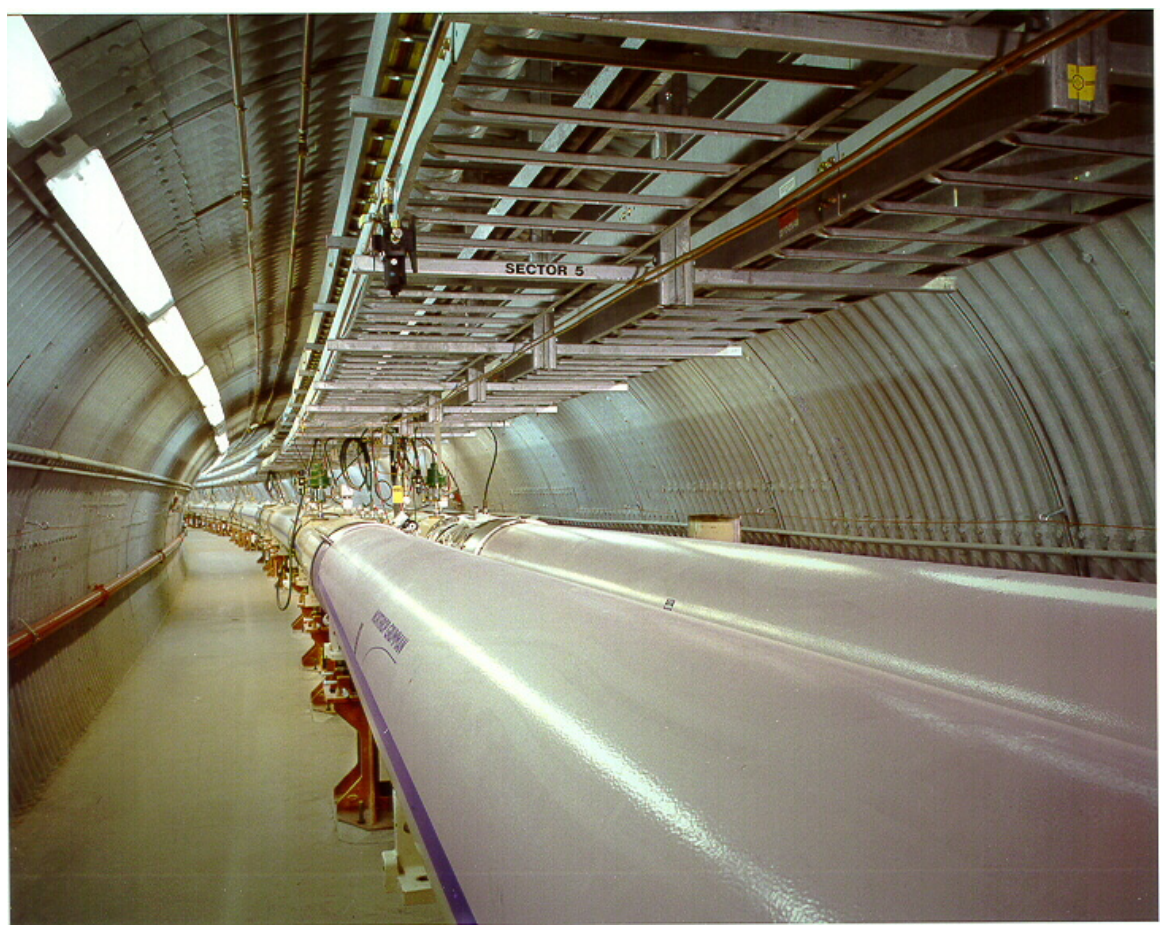

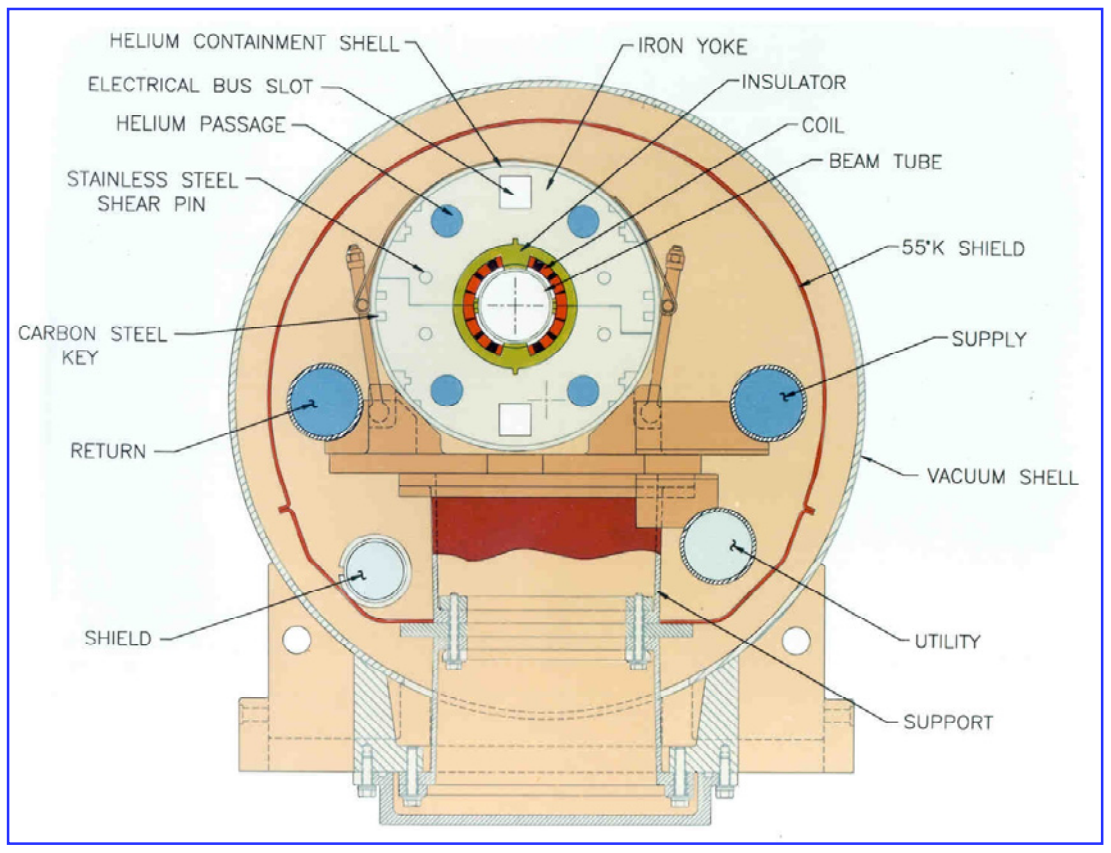

Figure 11. (left) Inside the RHIC tunnel—two rings of superconducting magnets. (right) Cross section of a RHIC dipole magnet viewed along the beam axis. B field is vertical. Note the resemblance to MRI magnet (Fig.7).

## 4.3 The New York City region nurtures science

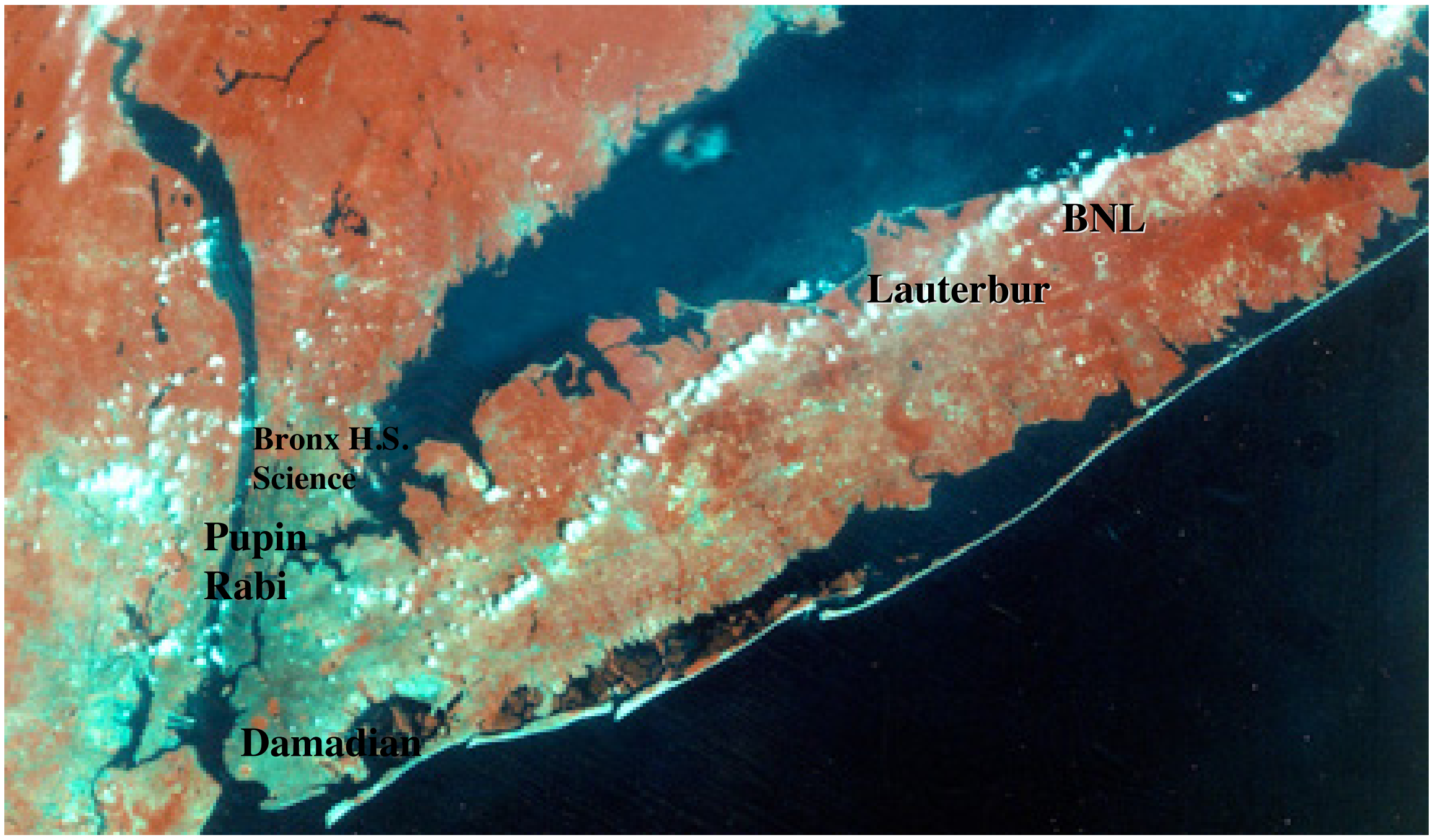

Figure 12. NASA Infra-red photo of New York Metro Region from Fig.9 with locations where work mentioned above was done. Also shown is the location of the Bronx High School of Science (see text). Not shown but also on this map are the original Bell Laboratories (in Manhattan until 1966), where the transistor was invented as well as many other discoveries, IBM Research Labs in Yorktown Heights, NY, Cold Spring Harbor Lab on Long Island, many research universities, etc.

Some locations where the fundamental science mentioned above was performed are shown (Fig.12) on the same map of the New York Metro region as Fig.9. I have also taken the liberty of showing the location of my High School, "Bronx Science" on the map since it is one of the great examples of nurturing science in the world.

## 4.4 The Bronx High School of Science counts seven Nobel Prize winning scientists among its graduates—all in Physics!

The Bronx High School of Science is a public high school (grades 9-12) in New York City open to all eligible students by competitive exam. No other secondary school in the world has as many alumni who have won Nobel Prizes. If Bronx Science were a country, it would be tied at 23rd with Spain for the number of Nobel Laureates (as of 2008). There are two other such H.S. in New York almost equally successful: Stuvesant H.S. in Manhattan and Brooklyn Technical H.S. Here is the list of Nobel Laureates (all in Physics) from Bronx Science:

- Leon N. Cooper 1947, Brown University, Nobel Prize 1972
- Sheldon L. Glashow 1950, Boston University, Nobel Prize 1979
- Steven Weinberg 1950, University of Texas at Austin, Nobel Prize 1979
- Melvin Schwartz 1949, Columbia University, Nobel Prize,1988
- Russell A. Hulse 1966, Princeton University, Nobel Prize 1993
- H. David Politzer 1966, California Institute of Technology, Nobel Prize 2004
- Roy J. Glauber 1941, Harvard University, Nobel Prize 2005

## 4.5 Many problems facing society at the beginning of the 21st Century need input from trained scientists.

- Climate change
- Clean renewable energy
- Nuclear power
- Nuclear proliferation
- ....

Perhaps more important, wide-spread scientific understanding in the general public is required in order to understand the validity of proposed solutions – or to help find the solutions! Thus in addition to specialized secondary school to produce trained scientists, the general science education in the public schools must be improved.

## 4.6 PHYSICS FIRST!—A Proposal to improve the science curriculum in U. S. secondary schools.

The standard science curriculum in a good U.S. secondary school is Biology, Chemistry, Physics. A movement led by the American Association of Physics Teachers (AAPT) and Leon Lederman of Fermilab is to revise the High School Curriculum to Physics First!

The three year coordinated science sequence would be Physics, Chemistry, Biology, while integrating Earth Science and Astronomy topics into these areas. The emphasis in a physics-first sequence should be focused on conceptual understanding rather than mathematical manipulation. Mathematics would be introduced on a "need-to-know" basis.

My elder daughter had a similar curriculum in High School and was very happy with it: Earth Science, Biology, Physics, Chemistry.

Physics First: : http://www.aapt.org/upload/phys_first.pdf
ARISE: http://ed.fnal.gov/arise/

# 5 The 21st century—beginning of the 3rd Millennium

The 20th century started with the study of macroscopic matter which led to the discovery of a whole new submicroscopic world of physics which totally changed our view of nature and led to new quantum applications, both fundamental and practical.

For the third millennium we start in the sub-nuclear world with a new periodic table to understand. Who can imagine where this will lead over the next century and beyond?

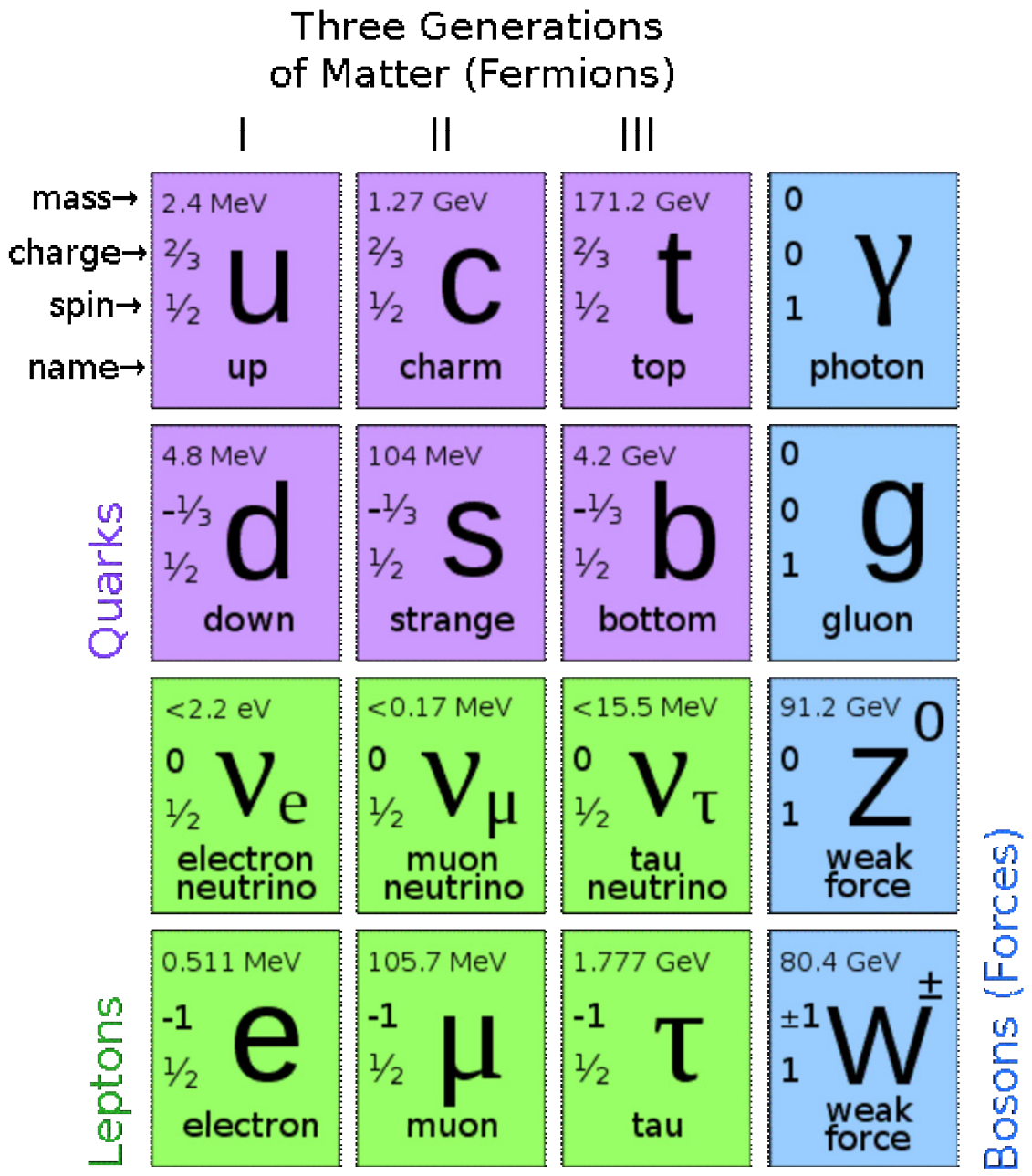

Figure 13. Circa 2001 understanding of the fundamental particles and forces of nature and the new set of submicroscopic quantum numbers, mass, charge, spin, quark, lepton, boson.